\documentstyle[times,pramana,psfig,floats]{ias}
\begin{document}
\markboth{Black Holes under External Influence}{J. Bi\v c\' ak}
\title{Black Holes under External Influence \footnote{The content
corresponds to the lecture given at ICGC 2000 in Kharagpur.
Sections 2-6 are based on the text of the lecture on
`Electromagnetic fields around black holes and Meissner effect'
given at the 3rd ICRA workshop in Pescara 1999 (to be published
with T. Ledvinka
in Nuovo Cimento).}}
\author{Ji\v r\' \i\ Bi\v c\' ak}
\address{Institute of Theoretical Physics,
Faculty of Mathematics and Physics,
Charles University\\
V Hole\v{s}ovi\v{c}k\'ach 2, 180 00 Prague 8,
Czech Republic}
\setcounter{page}{481}

\abstract{
The work on black holes immersed in external fields is reviewed
in both test-field approximation and within
exact solutions. In particular we pay attention to the effect of
the expulsion of the flux of external fields across charged and
rotating black holes which are approaching extremal states.
Recently this effect has been shown to occur for black hole
solutions in string theory. We also discuss black holes surrounded
by rings and disks and rotating black holes accelerated by strings.
}

\maketitle

\section*{Prologue}

Since the primary purpose of all physical theory is rooted in
reality, I shall start by an experiment. It is concerned with the
first terrestrial demonstration of a black hole, as observed in
Prague in 1913 and described by Gustav Meyring (1868-1932) in his
story `Black sphere'.

Two Indians came to central Europe to
give intriguing performances with a glass vessel to the throat of
which there was connected a small
golden chain. When the loose end of the chain
was attached to the forehead of a person, there appeared a
plastic picture inside the vessel which corresponded to the ideas
and images of the person. When one of the Indians joined the
chain to his forehead, a beautiful Indian landscape with Taj Mahal
appeared in the vessel. The clearest pictures arose from
the images of mathematicians. Strange and chaotic (`like Italian
salad') pictures corresponded to the ideas of lawyers and
psychiatrists... Once a not-quite - the cleverest officer took part
in the performance. When one of the Indians attached the golden
chain to the officer's forehead, nothing was happening for a long
time.

{Then, suddenly, there was a black sphere hovering freely
in the glass vessel. The astonished brahman took the vessel.
However, as he moved it, the sphere hovering inside touched the
glass wall. At the very same moment, the glass broke and the
splinters, as if being attracted by a magnet} [no, gravity!],
{have fallen into the sphere and disappeared there without leaving a
trace} [indeed, no-hair theorem, multipole moments get radiated out!].
{The black round body was hovering in free space. In fact, this
thing did not look as a sphere, rather, it looked as a black
hole} [!]. {It was something absolute - a mathematical
`nothingness'. What was happening next was only a necessary
consequence of the `nothingness'. Everything in a neighborhood of
the nothingness had, as a consequence of a natural necessity}
[GR], {fallen inside and changed immediately also into the
nothingness, i.e., it disappeared without a trace.}

{Indeed, suddenly a powerful stream of air arose as the air was
sucked into the hole. Pieces of paper, ladies' gloves and veils
were dragged inside ... the only people who remained were
the two Indians. `The whole Universe} [!]  {which Brahma created,
Vishnu keeps going and Shiva destroys, will subsequently
fall into  this sphere', said
R\'ad\v{z}endral\'alamitra solemnly - this, my brother, is a curse
due to our journey to the west! `What does it matter', murmered hosain,
`once we all must go into the realm of negative being anyway.'}

\section{Introduction}

Although much insight into the structure of external
fields around black holes was obtained in 1970s
and 1980s already, an interest in this theme continues, among
others also because of the
appearance of new motivations. To those coming from classical
relativity belongs, for example, the `membrane paradigm' [1] -- in the
following (Section 4) we shall briefly discuss its validity for almost
extreme black holes. Another `classical' issue is the behaviour
of fields on the Cauchy (inner) horizons of charged or rotating
holes [2]. Astrophysically, the discovery of microquasars in
our Galaxy [3] makes the mechanism of the energy extraction
from rotating black holes testable more directly than it has been
possible with supermassive black holes in distant galactic
nuclei. The role of the Blandford-Znajek mechanism [1] of
electromagnetic extraction of rotational energy is still not
properly understood. In Section 3 we shall discuss the flux of
stationary magnetic fields across rotating black holes. Some flux
across the horizon is needed in order that the Blandford-Znajek
mechanism may operate. First,
however, we shall briefly review our earlier work on coupled
perturbations of Reissner-Nordstr\"om black holes (Section 2).
In Section 5 exact spacetimes with black holes in strong
electromagnetic fields will be
considered. We shall here also mention the possibility of the existence of a
`cosmic supercollider' formed by a supermassive black hole
surrounded by a superstrong magnetic field [4].

New motivations have appeared with studies of black hole solutions in
spacetimes with the dimensions either lower or higher than four
within the framework of various `model' or `unified theories'. A
useful survey of these `generalized' black holes, in fact of
essentially all important developments in physics of black holes
up to 1998 is contained in the new monograph by V. Frolov and I.
Novikov [5]. In Section 6 we shall outline some new results on the expulsion
of magnetic (or more general) gauge fields from extremal black
holes in string theory.

Finally, in Sections 7 and 8 some most recent results obtained
in our Prague group will be mentioned. They are concerned with
exact spacetimes describing a Schwarzschild black hole surrounded
by a ring or disk of matter; and with the spinning $C$-metric,
which represents two uniformly accelerated, rotating black holes.
The cause of the acceleration is an `external' string between the
holes or strings extending from each of the hole to infinity.

\section{Perturbations of charged black holes}

The study of the perturbations of charged, non-rotating black
holes can be motivated from several viewpoints: (i) the
Reissner-Nordstr\"{o}m solution is the simplest solution
involving a limit - when the charge $Q$ and mass $M$ satisfy the
condition of extremality $(Q^2= M^2)$ - beyond which the horizon
is absent; (ii) the holes' interiors contain the Cauchy horizons
where generic perturbations should imply instability; (iii) the
electromagnetic perturbations are in general coupled with the
gravitational perturbations. This leads to such intriguing
effects as the conversion of gravitational waves into
electromagnetic ones, or the appearance of closed magnetic field
lines caused by `effective gravitational currents'; the
coupling can also be used to study rigorously the motion of a black
hole under an external, non-gravitational influence. Some of these
effects will be discussed in the following.

The necessity of the coupling can easily be understood from the
perturbed Einstein-Maxwell equations written symbolically in the
form

\begin{equation}
\delta G = 8\pi \delta T,
\label{1}
\end{equation}
where the perturbed energy-momentum tensor, $\delta T \sim F^{(0)}
\delta F$, is {\it linear} in perturbations because of the
presence of the background Coulomb-type electric field $F^{(0)} \sim
Q/r^2$. Hence, the left-hand side of (\ref{1}), the perturbed
Einstein tensor, is linear in metric perturbations $\delta g$ (and
their derivatives). The perturbed Maxwell equations, $F_{\alpha
\sigma}^{;\sigma}= 4\pi j_{\alpha}$, turn out to read explicitly
as follows [6]:

\begin{equation}
(\delta F_{\alpha \sigma })^{; \sigma} = 4\pi \delta j_{\alpha} +
(\delta j_{\alpha})_{grav.},
\label{2}
\end{equation}
where the effective `gravitational current' is given by

\begin{equation}
\left(\delta j_{\alpha}\right) _{grav.} = F^{(0)}_{\alpha \rho; \sigma}
h^{\rho \sigma} + F^{(0)\rho \sigma} h_{\alpha \rho ; \sigma}+
{{F^{(0)}}_{\alpha}}^{\rho} \left( h^{\sigma}_{\rho;\sigma}  -\frac{1}{2}h^{\sigma}_{\sigma; \rho} \right),
\label{3}
\end{equation}
where $h_{\rho \sigma} \equiv \delta g_{\rho \sigma}$. Since in
perturbed Einstein's equations (\ref{1}) and Maxwell's equations
(\ref{2}), (\ref{3}) the electromagnetic and gravitational
perturbations are coupled, it was not an easy task to convert the
formalism into a tractable form. We shall now very briefly sketch
the history of this issue, referring to the monographs [7],
[5] and to our extensive work [8] and review
[9] for details and citations of the original papers. A few recent
references will be given below.

The basic theory of interacting perturbations of the
Reissner-Nordstr\"{o}m black hole was started by Zerilli in
1974 who extended the Regge-Wheeler-Vishveshwara-Zerilli theory
of perturbations of the Schwarzschild black hole. After
decomposing perturbations into the (vector and tensor) harmonics,
Zerilli chose a coordinate system according to the Regge-Wheeler
gauge conditions. In this gauge he reduced the equations for
perturbations for each multipole with $l \geq 2$ to two equations
of the second order for two functions, the knowledge of which is
sufficient to determine all perturbations. Sibgatullin and
Alekseev, using a different gauge, found a pair of decoupled wave
equations in case of both parities. A novel approach to
investigate coupled perturbations was developed by Moncrief who,
by employing the Hamiltonian formalism, was able to find gauge
independent canonical variables in terms of which all metric and
electromagnetic perturbations can be determined after a gauge is
specified. The possibility to fix the gauge only towards the end
of calculations is advantageous not only from a principle,
`theoretical' point of view. In [10] we were able to find
explicitly all perturbations in the problem of the motion of a
charged black hole in an asymptotically uniform weak electric
field only because we chose a gauge different from the
Regge-Wheeler gauge. This choice was done after the equations for
gauge-independent quantities were solved. Moncrief also indicated
how the pair of decoupled wave equations can be obtained for
suitable combinations of gauge independent variables for
multipoles with $l \geq 2$.

In our extensive paper [8] we developed the formalisms for the
interacting perturbations of the Reissner-Nordstr\"om black holes
in detail and clarified the relations between them. First we
expanded Moncrief's theory: all Hamilton equations following
from four Moncrief's Hamiltonians (for $l \geq 2,~ l = 1$, and
in both parities) are derived in suitable forms and from them the
wave equations for gauge invariant perturbations are obtained.
Starting then from the Hamilton equations and employing the
relations between the standard form of perturbations, $\delta
g_{\mu \nu}$ and $\delta F_{\mu \nu}$, and canonical variables,
one can express all $\delta g_{\mu \nu}$'s and $\delta F_{\mu
\nu}$'s in terms of gauge invariant variables (satisfying the
wave equations) in a suitable gauge. In [8] this is done in
Regge-Wheeler gauge for $l\geq 2$ perturbations and in another
suitable gauge for the dipole perturbations. These results
enabled us to treat various perturbation problems (as, for
example, the expulsion of the field lines illustrated in Figure
1) and also to establish a detailed relation between the standard
(Zerilli-type) perturbation formalism and the canonical
(Moncrief-type) treatment of perturbations.

Of course, the coupled perturbations of the Reissner-Nordstr\"om
black holes were analyzed also in the framework of the
Newman-Penrose formalism which proved to be so efficient in the
case of rotating black holes. Lun, Chitre, Lee and Chandrasekhar
(see [7], [8] or [9] for references) are among
the main contributors to the theory. Their results are extended
in [8] in the following points: (i) the fundamental
perturbation variables which satisfy decoupled equations are not
only coordinate gauge invariant but also invariant under
infinitesimal transformations of the Newman-Penrose tetrads; (ii) the
dipole perturbations are analyzed in the Newman-Penrose formalism
for the first time, and they are treated simultaneously with the
$l\geq 2$ perturbations; (iii) the relations between the
canonical and the Newman-Penrose basic quantities are
established.

We are recalling these results not only because they form the
theoretical framework for solving the problems like the motion of
a charged black hole in an asymptotically uniform electric field
[10], or the fields of stationary sources on the
Reissner-Nordstr\"om background [6]. A somewhat
`central-European' character of the journal in which [8]
appeared, brings its ``fruits": in 1995 the formalism for the dipole
odd-parity perturbations of the Reissner-Nordstr\"om solution was
redeveloped in [11] and in 1999 the even parity case was treated
[12] without a realization that this was done in [8]
within the framework of three different formalisms. We
believe that in [8] the most complete discussion is given
of the interconnections between the standard, the canonical and
the Newman-Penrose formalisms even for perturbations of
Schwarzschild black holes.

Now there exist simple exact stationary multipole solutions for coupled
perturbations of the Reissner-Nordstr\"om black holes [13].
(Some of these solutions have very recently  been employed to study
the electromagnetic Thirring effects [14].) In [6] we
used these solutions to construct the magnetic field of a current
loop (magnetic dipole) placed axisymmetrically on the polar axis
of the extreme Reissner-Nordstr\"om black hole. The
electromagnetic and gravitational field occurring when the
general Reissner-Nordstr\"om black hole is placed in an
asymptotically uniform magnetic field was also derived.

\begin{figure}[t]
\centerline{\psfig{file=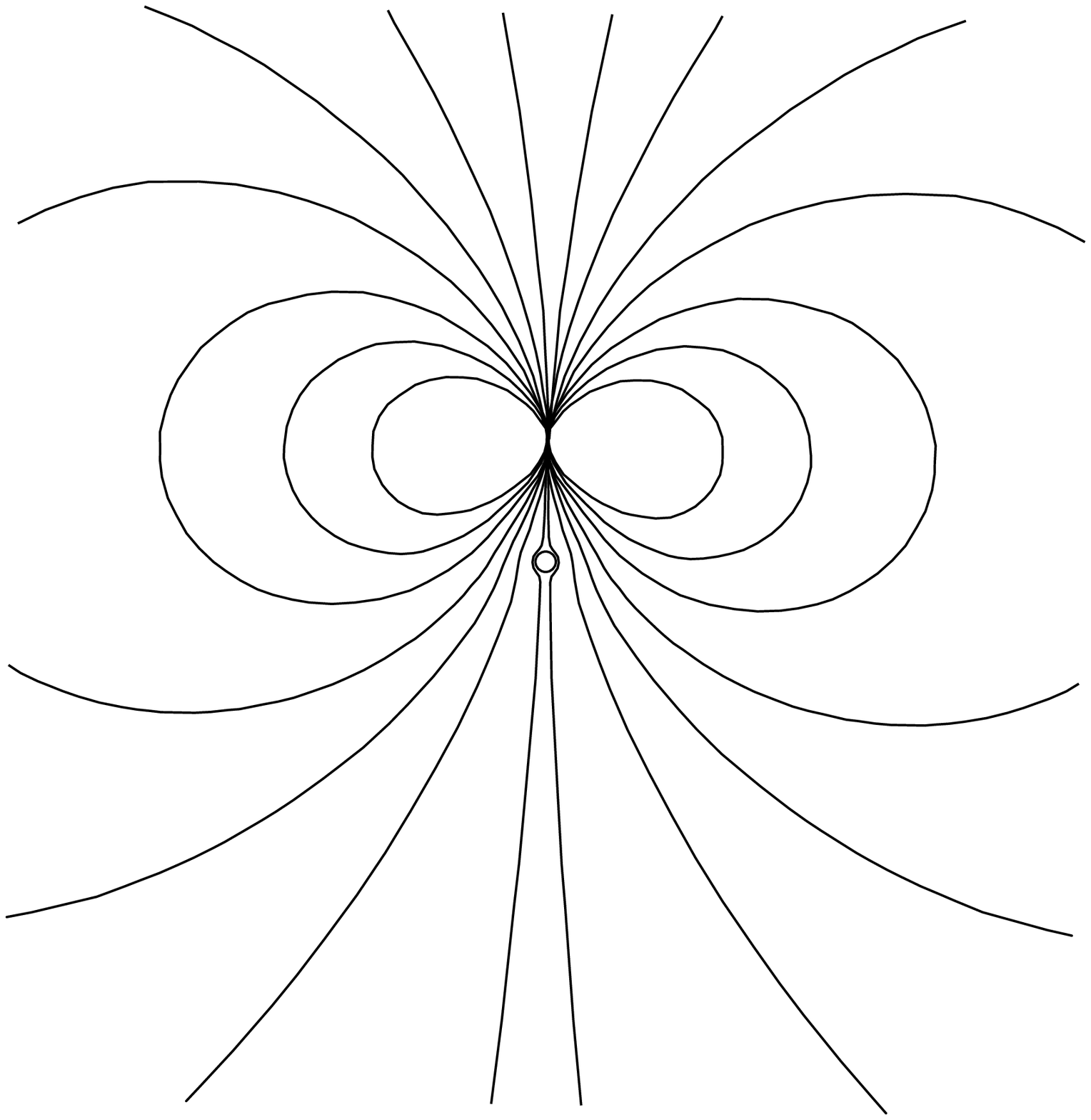,width=2.2in}\psfig{file=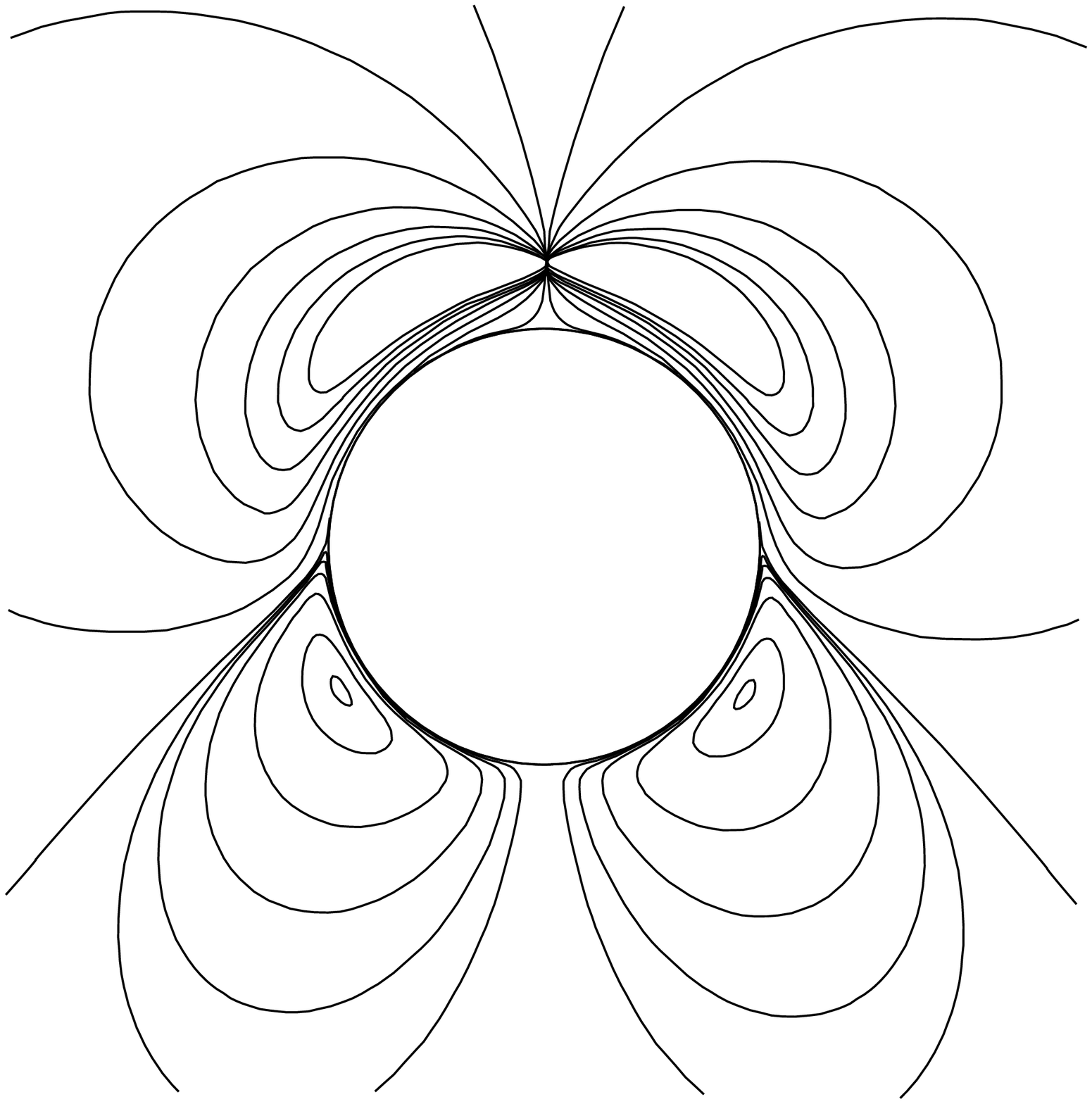,width=2in}}
\caption{
Field lines of the magnetic dipole placed (a) far away from
the extreme Reissner-Nordstr\"{o}m black hole, (b) close to the hole.}
\label{Fig1}
\end{figure}

The magnetic lines of force, as introduced by Christodoulou and Ruffini
(see [15] and references therein), were constructed
numerically for all the sources mentioned above at various
positions. We refer to [6] for details. Here, as an
illustration, we present Figure 1 (From [6]), showing magnetic lines of
force of the small current loop (magnetic dipole) located
axisymmetrically on the polar axis of the extreme
Reissner-Nordstr\"om black hole. One can make sure that
the structure of the closed lines of force in the region `opposite' to
the place where the magnetic dipole is located (Figure \ref{Fig1}b), is caused by the
coupling of electromagnetic and gravitational perturbations.
Owing to the extreme character of the black hole, no line of
force crosses the horizon.

\section{The flux of stationary magnetic fields across rotating black holes}
In order to investigate the structure of an asymptotically
uniform test magnetic field on the background of a Kerr black hole we
can start from the field $F_{\mu \nu}$ given explicitly in
[16]. Without repeating here complicated formulas, let us
recall that each $F_{\mu \nu}$ can be expressed as a sum of two
terms, one being proportional to $B_0$, the magnitude of the
component of the field asymptotically aligned with the hole's
rotation axis, the other, $B_1$, being the magnitude of the
component  perpendicular to the axis. Following Christodoulou and
Ruffini [15] we define the magnetic (electric) lines of
force as the lines tangent to the direction of the Lorentz force
experienced by a test magnetic (electric) charge at rest with
respect to the locally non-rotating frame. For the magnetic field
lines this definition yields $dr/d\theta = -F_{\theta \phi}/F_{r
\phi} \equiv B_r/B_\theta$ and $dr/d\phi = F_{\theta \phi} / F_{r
\theta} \equiv B_r/B_\phi$.

\begin{figure}[h]
\centerline{\psfig{file=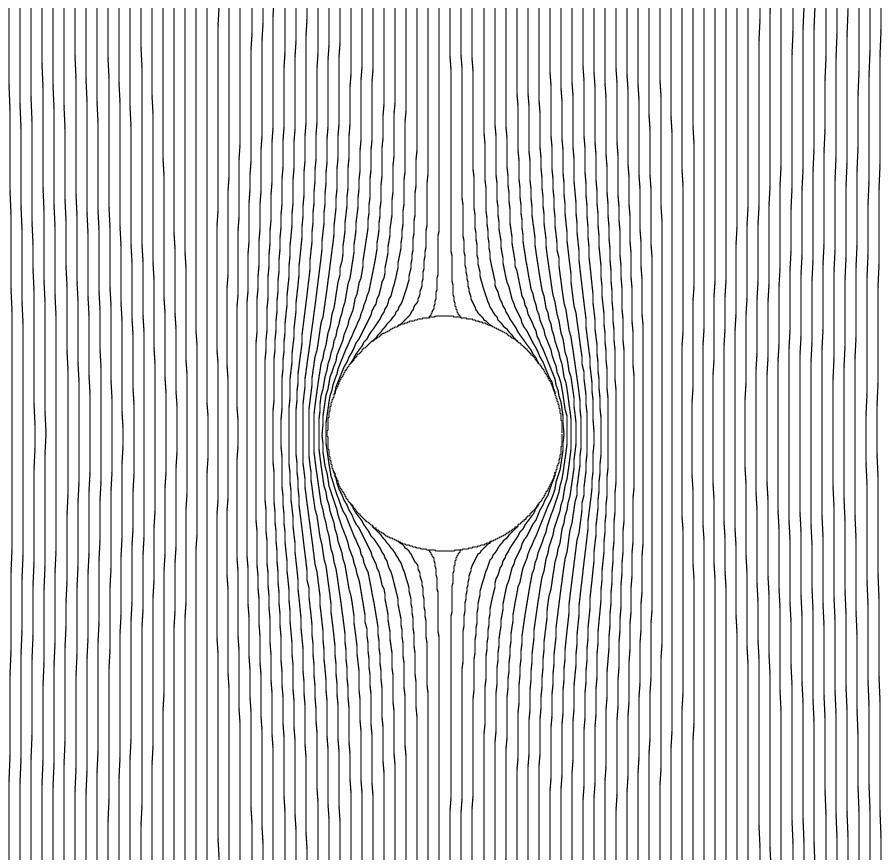,height=2.3in}~\hfil\psfig{file=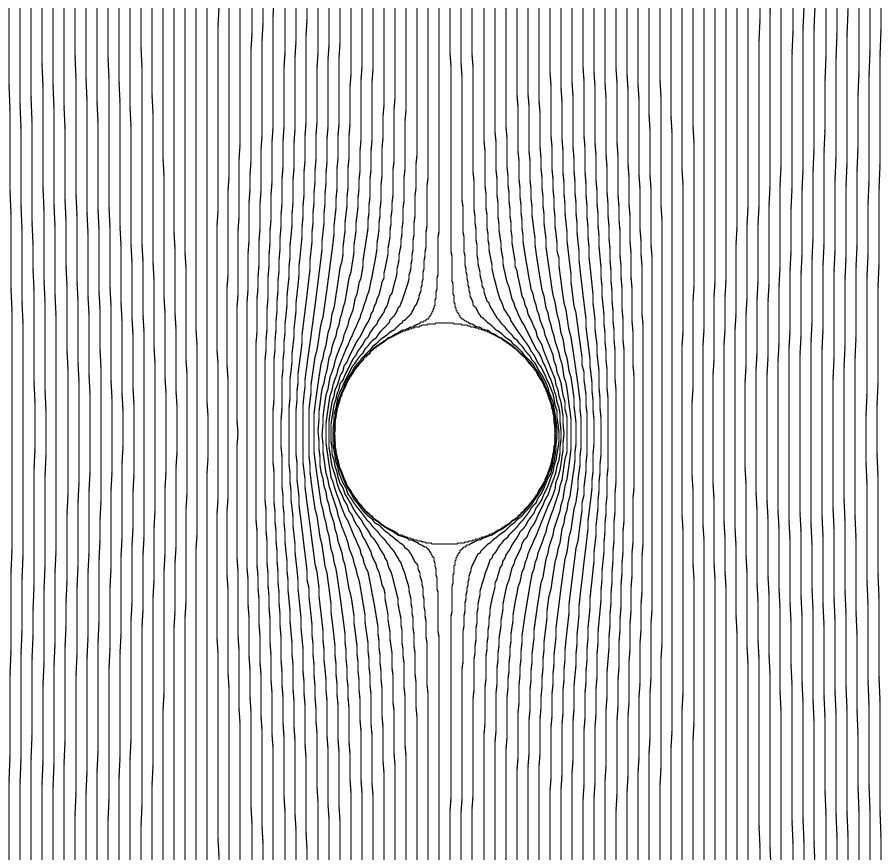,height=2.3in}}
\centerline{(a) \hskip 6cm (b)}
\caption{
Field lines of the test magnetic field homogeneous at infinity and aligned with hole's rotation axis.
The cases with (a) $a=0.998 M$ and (b) $a=M$ are shown.}
\label{Fig3}
\end{figure}

In the case of the aligned field we can easily verify (by using
$F_{\mu \nu}$ from [16]) that the field lines lie on the
surfaces of constant flux,

\begin{equation}
\phi = \pi B_0 [\Delta + 2Mr \Sigma^{-1} \left( r^2 - a^2 \right) ]
\sin ^2 \theta = {\rm constant},
\label{9}
\end{equation}
where

\begin{equation}
\Delta = r^2 - 2Mr + a^2, \,\,\, \Sigma = r^2 + a^2 \cos ^2 \theta;
\label{10}
\end{equation}
$r, \theta, \phi$ are Boyer-Lindquist coordinates. The field
lines constructed numerically are shown in Figure \ref{Fig3}. The figures
clearly illustrate how the magnetic field is expelled from the horizon
when the angular momentum of the hole increases. Analogously to
the Reissner-Nordstr\"om case, no field line of the
asymptotically uniform magnetic field enters the horizon of an
extreme Kerr black hole.

\begin{figure}[ht]
\centerline{\psfig{file=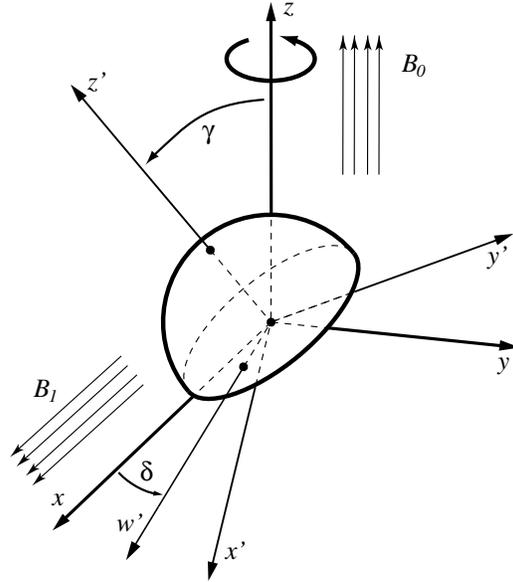,width=2.7in}}
\label{Fig4}
\caption{
A generally located hemisphere across which the magnetic flux of an asympotically uniform magnetic field is calculated.
}
\end{figure}

In the case of {\it general} stationary axisymmetric (i.e. ``aligned")
electromagnetic fields, one arives, using solutions given in [17]
(in the Newman-Penrose formalism), at a similar result:
the flux of an arbitrary, axially symmetric stationary magnetic field across any part of the
horizon of an extreme Kerr black hole vanishes.

The structure of an asymptotically non-aligned field is much more
complicated. In this case $B_\phi \not= 0$ and the field lines are
dragged around the black hole. The lines, originally parallel to
each other, are twisted, some of them threading the horizon even
in an extreme case. The results of the numerical construction of
the magnetic lines asymptotically perpendicular to the rotation
axis of the Kerr black hole with $a/M=0.998$, as they look in the
equatorial plane when viewed from above, are given in [18]. The
field lines in Boyer-Lindquist coordinates are wound up around
the horizon; in the Kerr ingoing coordinates the field lines do not wind up.

The structure of the magnetic field near a black hole can be
characterized by the magnetic flux across (half of) the horizon.
A general position of the hemisphere can be specified on the
basis of an Euclidean picture. The magnetic flux across the
hemisphere can be defined invariantly as an integral over the
surface of the horizon [16]

\begin{equation}
\Phi = \int^{2\pi}_{0} \!\!\!\!\! d \phi \int^{\Theta(\phi;\gamma ,
\delta)}_{0} \!\!\!\!\!\!\!\!\!\!\!\!d\theta~{F_{\theta \phi}}_{|r=r_+},
\label{14}
\end{equation}
where
$\Theta (\phi ; \gamma , \delta) \equiv \frac{\pi}{2} + \arctan
[\tan \gamma \cos (\phi-\delta)] $ - see Figure \ref{Fig4}.

Substituting for $F_{\theta \phi}$ and performing the integral we find rather complicated expression which,
however, can be understood intuitively in special cases: (i) If $\gamma=0$ the magnetic flux reduces to
$ \Phi = B_0  \pi r_+^2 (1-a^4/r_+^4)$; with $\gamma=0$ the total flux of the $B_1$ component of course vanishes.
As the hole becomes extreme we find $\Phi \rightarrow 0$ -- see Figure \ref{Fig3} (c-d).
(ii) If $\gamma = \pi/2$ the flux is $ \Phi = B_1 \pi[r_+^2 \cos \delta - (r_+ + M)a\sin \delta]$.
If the hole is not rotating one gets $ \Phi = B_1 \pi r_+^2 \cos \delta = \Phi_{\rm classical}$.
We get zero flux for $\delta=\pi/2$ and maximal one for $\delta = 0$. When the hole rotates the field lines are dragged along
and the flux vanishes across the hemisphere rotated by an angle $(\pi/2)-\delta_0$ where $\delta_0 \sim 27^o$ for $a=M>0$.
For a given $a$ there exists $\delta_{\rm max}$ for which the flux is maximal. With $a=M$ we find $\delta_{\rm max} \sim -63^o$,
$\Phi \sim 2.25 B_1 \pi r_+^2 = 2.25 \Phi_{\rm classical}$.
Hence, we see that although the flux of the aligned component $B_0$ decreases to zero with $a \rightarrow M$, the flux of the
perpendicular component is enhanced.

\section{Magnetic flux across the stretched horizon of an almost
extreme Kerr black hole}

In this section we shall outline some results of [19]
and of the discussions one of us (J.B.) had with Richard
Price in 1996. The discussions were concerned with the validity of
the membrane paradigm [1] for `alex' black holes, as we
call `almost extreme' black holes. It is well-known that a proper
distance from any point outside an extreme black hole to the outer
horizon is infinite. This fact implied the statement in
[1], p. 120, that `because the horizon is infinitely far
down in the [embedding] diagram, the finite magnetic flux has an
infinite spatial distance in which to wrap itself around the
embedding cylinder, and, consequently, near the horizon the
magnetic field falls off to zero.'

Notice, first, that in fact in a freely falling frame the
component $B_{(\theta)}$ and $B_{(\varphi)} = -B_{(\theta)}\cos \theta$ do
not vanish at the horizon of an extreme Kerr black hole, only the
radial component $B_{(r)}= 0$. (In case of an extreme
Reissner-Nordstr\"{o}m black hole all components of magnetic
field do vanish even in a freely falling frame at the horizon.)
Now in the spirit of the `membrane paradigm' consider a flux
across a stretched horizon characterized (in the Boyer-Lindquist
coordinates) by the redshift factor for locally non-rotating
frames (`lapse function'),

\begin{equation}
\alpha = \left[ (r^2 + a^2 \cos ^2 \theta) \Delta \right]
^{\frac{1}{2}} \left[ ( r^2 + a^2)^2 - a^2 \Delta \sin^2 \theta
\right] ^{-\frac{1}{2}},
\label{4}
\end{equation}
which at a stretched horizon is small, $\alpha_{Hs} = {\rm constant}
\ll 1$, but nonvanishing - in contrast to the true horizon for
which $\alpha_{Ht} = 0$. Introduce a parameter $\epsilon =
1-a/M$ so that $\epsilon$ is small for alex black holes and
vanishes for the extreme holes. We know that the magnetic flux of
axisymmetric fields vanishes at the true horizon in the limit $\epsilon
\to 0$:
\begin{equation}
\lim_{\epsilon \to 0} \phi_{|\alpha_{Ht}=0} = 0.
\end{equation}
A question arises whether if we `exchange' the limits, i.e.,
calculate first the flux across the stretched horizon of an alex
black hole and then go to the extreme case we could obtain a
nonvanishing quantity: will then

\begin{equation}
\lim_{\alpha_{Hs} \rightarrow 0} \left(~
   \lim_{\epsilon \rightarrow 0}
        {\phi_{|\alpha_{H_s\not=0}}} ~\right) \not= 0~?
\label{5}
\end{equation}
Restricting ourselves to asymptotically uniform fields we can
easily calculate the flux across the stretched horizon (using the
expressions for electromagnetic field given in [16]) and
find that

\begin{equation}
\Phi \approx 8 \pi B_0 M^2 \alpha_H ~~ {\rm for} ~~
\epsilon \ll \alpha^2_H~,~~~~
\Phi \approx 4\pi B_0 M^2 (2\epsilon)^{\frac{1}{2}} ~~{\rm for}~~ \epsilon
\gg \alpha^2_H\,.
\label{6}
\end{equation}
Therefore, $\Phi$ across the stretched horizon is nonvanishing
even in the extreme case $\epsilon = 0$ but it depends on where
$H_s$ is located; however, $\Phi \rightarrow 0$ as $\alpha _{Hs}
\rightarrow 0$, so that the limit (\ref{5}) vanishes. For any $\epsilon
>0$ there exists $\delta > 0$ such that $\alpha_H<\delta$ and
$\Phi < \epsilon$. This conclusion seems to suggest that power in
the Blandford-Znajek model arises from regions with `relatively
large' $\alpha _H$ in near-extreme case.

\section{Black holes in magnetic fields: exact models}
We now turn to the exact stationary solutions of the
Einstein-Maxwell equations representing rotating, charged black
holes immersed in an axisymmetric magnetic field. In the
weak-field limit -- when $\beta \equiv B_0 M \ll 1$, the constant
$B_0$ in the weak-field limit characterizes the magnetic field
strength, $M$ the hole's mass -- there exists a region
$2M\ll r\ll {B_0}^{-1}$ where the spacetimes are approximately flat and
the magnetic field is approximately uniform. At $r\gg B_0^{-1}$ the
metrics approach Melvin's magnetic universe.

The simplest example is the magnetized Schwarzschild black hole:

\begin{equation}
ds^2 = \Lambda ^2 \left\{{ dr^2\over 1-{2M\over r}} + r^2 d \theta ^2 -
(1-{2M\over r})dt^2\right\} + \Lambda ^{-2} r^2 \sin ^2 \theta \,\, d \phi^2,
\label{11}
\end{equation}
where $\Lambda = 1 + \frac{1}{4} B^2_0 r^2 \sin ^2 \theta$, the
electromagnetic field being given by $F_{\theta \phi} \equiv B_r = B_0
\Lambda ^{-2} r^2 \sin \theta \cos \theta$, $F_{r \phi} \equiv
-B_\theta = B_0 \Lambda^{-2} r \sin ^2 \theta$. The magnetic flux
across a general hemisphere on the horizon (with $\delta = 0$ due
to axisymmetry, cf. Figure \ref{Fig4}) reads as follows:

\begin{equation}
\Phi = 4 \pi \beta M \left[ 1+\beta ^2 \right] ^{-\frac{1}{2}}
\left[ 1+\beta ^2 + \tan ^2 \gamma \right] ^{-\frac{1}{2}}.
\label{12}
\end{equation}
For the flux across the `upper half' of the horizon $(\gamma =
0)$, we obtain the result given in [16]. (See Eq. (41) therein where
incorrect $M$ should be replaced by $r_+ = 2M$ in the
denominator.)

If $\beta \ll  1$ the flux is equal to $4\pi \beta M (1+\tan ^2
\gamma)^{-\frac{1}{2}}$ in accordance with the weak-field
approximation. By increasing $\beta$ (with $M$ and $\gamma$ fixed)
$\Phi$ first increases, as we expect on classical grounds.
However, a further increase of $\beta$ leads to the decrease of
the flux. Indeed, there exists a value of the magnetic field
parameter $\beta$ for which the flux acquires its maximum value,
$\beta = (1+\tan ^2 \gamma)^{\frac{1}{4}}$. The global maximum,
occurring
with $\gamma = 0$, is
\begin{equation}
\Phi=\Phi_{max}\equiv 2\pi M \doteq 3.22 \times 10^{39}
(M/10^9M _{\odot}) \left[ {\rm Gauss\;cm^2} \right].
\label{R13}
\end{equation}
The existence of the limiting value of the magnetic flux across
the horizon is caused by the gravitational effect of the magnetic field
which concentrates itself near the axis of symmetry when $\beta \gg 1$.

In [4], a rather exotic but interesting model was presented recently
in which a supermassive black hole is surrounded by a superstrong magnetic field
of the strength corresponding to just the half of the maximal flux (\ref{R13}).
In this model an electric induction field can accelerate charged particles up to an energy
$10^{18}$ GeV.

There exists a fairly extensive literature on exact solutions representing
rotating charged black holes in an external magnetic field (a magnetic Melvin-type universe).
They can be used to study the Meissner effect within the exact framework. One can make sure
that the magnetic fluxes vanish across the horizons of extreme black holes,
i.e. those with zero surface gravity. We refer to [20] and to very recent paper [21] for the review
and relevant references.

\section{Meissner effect for superconducting branes and extremal black
holes in string theory}

In 1998 a comprehensive paper by Chamblin, Emparan and Gibbons
[22] reviewed and developed evidence for the Meissner
effect for extremal black hole solutions in string theory and
Kaluza-Klein theory. Here we shall mention some of their ideas
and results, sticking closely to their discussion. They first
give a brief phenomenological description of the Meissner effect
in superconductors. Using London equation
\begin{eqnarray}
\vec{j} = - \,{\rm constant}~ \vec{A}~,
\label{7}
\end{eqnarray}
($\vec{j}$ - current, $~\vec{A}$ - vector potential),
and Maxwell's equations, one arrives at the equation
\begin{equation}
\nabla ^2 \vec{A} = \lambda ^2 \vec{A},
\label{8}
\end{equation}
$\lambda = {\rm constant}$, given by constants characterizing the
material. The solution in one dimension, $A \sim \exp(-\lambda
x)$, indicates that $\vec{A}$ (and thus the magnetic
field as well) decreases exponentially as one goes from the
surface into the superconductive material. This is the classical
Meissner effect. We see that the expulsion of a magnetic field
from extreme (standard) black holes is analogous -- but there is a
difference: no magnetic flux at all penetrates into a hole. The
relativistic generalization of the London equation reads

\begin{equation}
J_{\mu} = - \lambda^{-2} A_{\mu} + \partial _{\mu}
\Lambda,
\label{RLON}
\end{equation}
or

\begin{equation}
-\lambda ^{-2} F_{\mu \nu} = \partial _{\mu} J_{\nu} - \partial
_{\nu} J_{\mu},
\label{RLON2}
\end{equation}
where $J_\mu$ is $4$-current and $A_\mu$ 4-potential.
In [22] equation (\ref{RLON2}) is adopted as the criterion
for superconductivity. The authors first review the work of
Nielsen and others showing that this equation is satisfied in the
Kaluza-Klein theory on the world volume of extended objects
carrying Kaluza-Klein currents. Then they consider self-gravitating
extended superconducting objects and demonstrate how the flux of
a gauge field is expelled from the intersection of two sets of
6-dimensional self-gravitating branes of 11-dimensional supergravity.

Several examples are then constructed in [22] demonstrating
that the expulsion of gauge fields occurs always at the extreme
horizon. Consider a string in $d=5$ which is wrapped along the
string direction so that a dilatonic black hole solution in $d=4$
arises. The starting metric in $d=5$ reads

\begin{equation}
ds^2 = H^{-\frac{1}{3}} \left( -f dt^2 + dz^2 \right) +
H^{\frac{2}{3}} \left[ f^{-1} dr^2 + r^2 \left( d\theta ^2 + \sin
^2 \theta d \varphi ^2 \right) \right],
\label{15}
\end{equation}
where
$H = 1+q/r$, $f=1-r_0/r$ and
$q, r_0$ are constants, an event horizon being at $r=r_0$;~ $r_0=0$
is the extreme case. If this geometry is compactified along the
string direction $z$ in such a way that the compactification
direction is twisted -- the compactification is made along the orbits of the vector
$(\partial/\partial z) + B(\partial/\partial\varphi)$, where
constant $B$ will describe the asymptotic value of the magnetic
field along the axis -- one obtains a black hole and the
Kaluza-Klein magnetic field. The field is described by the
potential whose $\varphi$ - component is given by

\begin{equation}
{\cal {A}} = B g_{\varphi \varphi} / (g_{zz} + B^2 g_{\varphi
\varphi}) = B Hr^2 \sin ^2 \theta / (1+B^2 Hr^2 \sin ^2 \theta).
\label{17}
\end{equation}
In general one finds a non-vanishing flux across the horizon at
$r=r_0$. In the limit of an extreme hole, however, at the horizon
$r=0$ the potential ${\cal{A}}$ vanishes and no magnetic flux
thus penetrates the horizon.

As another example of the Meissner effect Chamblin, Emparan and
Gibbons consider the field expulsion from extreme rotating black
holes. First they recall the expulsion of the asymptotic uniform
magnetic test field in the extreme Kerr black hole background
discussed in Section 3. Then they construct an exact solution (by
taking the product of the standard Kerr solution with
$x^5$-direction and applying a twisted reduction procedure,
similar to that considered above but involving also a twist in
the time coordinate) representing a Kerr black hole in the {\it
exact} Kaluza-Klein gauge field. Again, this exact field exhibits
the Meissner effect in the extreme case.

At the end of the subsection (see [22], {\it V.} A.) the authors
write that they considered the solutions in which the magnetic
field is aligned with the rotation axis of the black hole but
that according to our work [6] the Meissner expulsion can
also be seen for fields where no alignment is assumed. This is
not correct: as mentioned in the previous Section 3, and
demonstrated in detail in [16], the asymptotically
non-aligned fields do penetrate the extreme Kerr horizons. Only
general {\it axisymmetric, stationary} fields on the Kerr
background exhibit the Meissner effect in the extreme limit. On
the other hand, it is well known that configurations with
stationary external fields which are not axisymmetric, are not
stable -- due to the torque exerted on the horizon by an external
non-axisymmetric field (see e.g. [1]) - and evolve towards
axisymmetric configurations.

In case of the standard superconductivity, when the Meissner
effect arises, the field inside a superconductor becomes a pure
gauge. This is not the case with extreme black holes. In
[6] in Figure 1(d) the field lines are constructed inside
the extreme Reissner-Nordstr\"om horizon. Clearly, one cannot
claim that the interior of extreme black holes is in a
superconducting state.

The question of the flux expulsion from the horizons of extreme
black holes in more general frameworks is not yet understood
properly. The authors of [22] `believe this to be a
generic phenomenon for black holes in theories with more
complicated field content, although a precise specification of
the dynamical situations where this effect is present seems to be
out of reach.' That for abelian Higgs vortices the phenomenon of
flux expulsion from extreme black holes does not occur in all
cases has been argued analytically and investigated numerically
recently [23]. In particular, it appears that thin
cosmic strings (modelled by the vortices) can pierce the extreme
horizons whereas a thicker string will be expelled.

\section{Black holes under the influence of external rings and disks:
exact models}

We shall now briefly pay an attention to the recent work of
Semer\'ak, Zellerin and \v{Z}\'a\v{c}ek [24], [25] in
which exact solutions are constructed, representing static
axisymmetric matter distributions around non-rotating black
holes. There exists an extensive literature on disks surrounding
black holes. The disks are usually treated as a test matter, or
within a perturbation theory, or by numerical techniques. We refer
to the Introduction in [24] for a number of citations to
relevant papers.

In new work [24] Semer\'ak et al used the `old' fact that
in case of static axisymmetric vacuum spacetimes Einstein's
equations reduce considerably: one equation becomes Laplace's
equation for one metric function, the other equation determines
the second function in terms of a quadrature. The linearity of
Laplace's equation enables one to find fields representing
`superposed sources'. In the standard Weyl coordinates the
Schwarzschild black hole with mass $M$ is well known to be
described by a rod of length $2M$ (cf. [26]) along the axis
of symmetry. In [24] this source, and also  the
Appell ring (the potential field generated by a particle
situated on the imaginary extension of the axis of symmetry),
is superposed with the following additional sources in the equatorial plane:
(i) the Bach-Weyl ring (a `standard' ring placed in the
equatorial plane), (ii) the annular disk. The authors of [24]
construct the complete fields and then plot nice figures
illustrating the `gravitational field lines' (defined as
integral curves of the acceleration fields) in all three cases
for various values of masses of the hole and the `external'
source. In another set of figures they illustrate the distortion
of the Schwarzschild horizon caused by the presence of the
external matter. In the second paper [25] a detailed
description is given of how the motion of test particles is
influenced by the superposed fields, in particular how the
equatorial geodesics are modified when, for example, the mass of
an external disk-like source is gradually increased.

\section{The spinning C-metric: an exact model of accelerating, rotating black holes}
The static part of a spacetime representing the standard, non-spinning vacuum C-metric was originally
found by Levi-Civita in 1917-1919  (see references in [26]) but it was only in 1970
when Kinnersley and Walker understood, by choosing  a better parametrization,
that it can be extended so that it represents two black holes uniformly accelerated
in opposite directions. The ``cause" of the acceleration is given by nodal (conical) singularities
(``strings" or ``struts") along the axis of symmetry. By adding an external gravitational
field the nodal singularities can be removed [27]. If the holes are electrically (magnetically) charged
the field causing the acceleration  can be electric  (magnetic) [28,10].  These types of
generalized C-metric have been recently used in the context of quantum gravity - to describe
production of black hole pairs  in strong background fields (see e.g. [29]).

From the 1980's  several other works analyzed the standard C-metric.
In none of these works, however, has  the basic fact   been emphasized that the
vacuum C-metric is just one specific  example  in a large class of asymptotically
flat radiative spacetimes with boost-rotation symmetry (with the boost along the axis of
rotational symmetry).
From a unified point of view,
boost-rotation symmetric spacetimes with hypersurface orthogonal axial and
boost Killing vectors were studied geometrically in [30].
We refer to this detailed work for rigorous definitions and theorems. In fact it is no surprise
that the C-metric was for a long time analyzed in coordinate systems  unsuitable for
treating global issues such as the properties of null infinity. It is algebraically
special, and the coordinate systems were adapted to its degenerate character.
In  polar coordinates $\{t,\ \rho,\ \phi,\ z\}$ the metric  of a general boost-rotation
symmetric spacetime with hypersurface orthogonal axial and boost Killing vectors,
${\partial \over \partial \phi}$ and $z{\partial\over\partial t} + t\frac{\partial}{\partial z}$,
has the form  (see [30], Eq. (3.38)):

\begin{eqnarray}
{\rm d}s^2 =&e& {}^{\lambda} {\rm d} \rho^2 + \rho^2 e^{-\mu} {\rm d} {\phi}^2\nonumber\\
 &+& \frac{1}{z^2-t^2} \left( (e^{\lambda} z^2- e^{\mu} t^2)      {\rm d}z^2 \right. \nonumber\\
 &-&  \left. 2zt(e^{\lambda}-e^{\mu}){\rm d}z {\rm d}t - (e^{\mu} z^2 -e^{\lambda} t^2 ){\rm d}t^2  \right) \ ,
\label{bsrotmA}
\end{eqnarray}
\vskip 3mm
where $\mu$ and $\lambda$ are functions of $\rho^2$ and $z^2-t^2$, satisfying, as a
consequence of Einstein's vacuum equations, a simple system of three equations, one
of them being the wave equation for the function $\mu$.
The whole structure of the group orbits in boost-rotation symmetric curved spacetimes  outside
the sources (or singularities) is the same as the structure of the orbits generated by the
axial  and boost Killing vectors in Minkowski space.  In particular, the boost Killing
vector  is timelike  in the region $z^2>t^2$.  It is this region which can
be transformed  to the static Weyl form.  Physically this corresponds to the transformation
to ``uniformly accelerated frames" in which sources are at rest and the fields are time independent.

However, in the other ``half" of the spacetime, $t^2>z^2$  (``above  the roof" in the terminology of
[30]),  the boost Killing vector is spacelike so that in this region the metric
(\ref{bsrotmA})  is nonstationary. It can be shown  that for $t^2>\rho^2+z^2$  the metric can be
locally  transformed into the metric  of the Einstein-Rosen waves. The  radiative properties
of the specific boost-rotation symmetric spacetime were investigated in a numer of papers; these
spacetimes  were also used as test beds
in numerical relativity - see the review [31] for more details and references.

Now in all the work mentioned above it was assumed that the axial and boost Killing vectors
are hypersurface orthogonal. Recently, we analyzed symmetries compatible with asymptotic flatness and
admitting gravitational and electromagnetic radiation [32]. We have shown that in
axially symmetric electrovacuum spacetimes in which at least locally a smooth null infinity
exists, the only second allowable  symmetry which admits  radiation  is the boost symmetry.
The axial and an additional Killing vector have {\it not}  been  assumed to be hypersurface
orthogonal.  In [32]  the general functional forms of gravitational  and electromagnetic
news functions, and of the total mass of asymptotically flat boost-rotation symmetric spacetimes
at null infinity have been obtained. However, until now no general theory  similar to that
given in [30]  in the hypersurface orthogonal case is available for the boost-rotation
symmetric  spacetimes with Killing vectors which are not hypersurface orthogonal.
 Nevertheless
there is one explicitly given metric which can be expected to serve as an example
of these spacetimes - the spinning vacuum C-metric.
It was discovered by Pleba\'nski and Demia\'nski [33]  in 1976.
However, its  boost-rotation character  has not been analyzed.  In particular, ``the canonical
coordinates"  in which the metric   represents
a generalization of the metric (\ref{bsrotmA}) so that global issues outside the
sources could properly
be studied have not been found so far.

In our recent work [34] we found such  a  representation of the spinning
C-metric (SC-metric)  which  generalizes (\ref{bsrotmA}) and, thus,  can
also serve as a convenient example  for building up the general theory of
boost-rotation symmetric spacetimes with the Killing vectors which are in general
not hypersurface orthogonal.

We first specialized  the Pleba\'nski-Demia\'nski class of metrics
[33] to the  spinning vacuum case, discussed the ranges of parameters
entering  the metric and indicated the limiting
procedure leading to the Kerr metric.  Then the SC-metric was transformed into Weyl
coordinates.
We showed that, by choosing  different values of the original Pleba\'nski-Demia\'nski coordinates we can,
in principle, arrive at various  Weyl spacetimes.
The properties of Killing vectors and of some invariants of the Riemann tensor
led us to choose a physically plausible Weyl spacetime
which contains  both the black-hole  and the acceleration horizon. Similarly as has been done
with  the standard  C-metric  [35],  we concentrated on this Weyl portion.
Remarkably, the metric  can then be also transformed into the ``canonical form"  of boost-rotation
symmetric spacetimes:

\begin{eqnarray}
{\rm ds}^2 = {\rm e}^{\lambda} {\rm d} \rho^2 + \rho^2 {\rm e}^{-\mu} {\rm d} \varphi^2
-    2A {\rm e}^{\mu} (z {\rm d}t -  t {\rm d} z)  {\rm d} \varphi -A^2 {\rm e}^{\mu} (z^2-t^2)   {\rm d} \varphi^2
\nonumber \\
+\frac{1}{z^2-t^2} \left[ ({\rm e}^{\lambda} z^2  - {\rm e}^{\mu} t^2 ) {\rm d} z^2 -   2zt ({\rm e}^{\lambda} - {\rm e}^{\mu}  ) {\rm d} z {\rm d} t +     ({\rm e}^{\lambda} t^2 - {\rm e}^{\mu} z^2 ) {\rm d} t^2  \right] \ ,
\label{BStvar}
\end{eqnarray}
\vskip 3mm
where
$\mu$, $\lambda$ and $A$ are functions of $\rho^2$ and $z^2 - t^2$, connected with the original
Pleba\'nski-Demia\'nski coordinates and with the Weyl coordinates in a quite complicated manner
(see [34] for details).

The form (\ref{BStvar}) represents a general boost-rotation symmetric spacetime
``with rotation", i.e.,  with the Killing vectors which are not hypersurface orthogonal.
Putting $A=0$, we recover the canonical form of the boost-rotation symmetric spacetimes with the
hypersurface orthogonal Killing vectors the general structure of which was studied in
[30] .

Analogous to the C-metric
without spin, the axis of symmetry contains nodal singularities
between the spinning ``sources" (holes)  which cause  the acceleration, and is regular  elsewhere;
or the axis can be
made regular between the sources but then the nodal  singularities  extend from
each of the sources to infinity (see Fig. 4 taken from [34]).

\begin{figure}[t]
\centerline{
\psfig{file=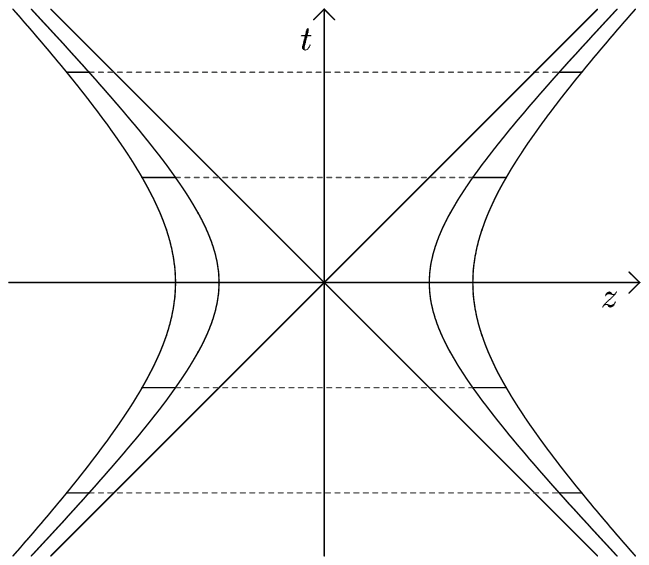,height=2in}
\hspace{0.2mm}
\psfig{file=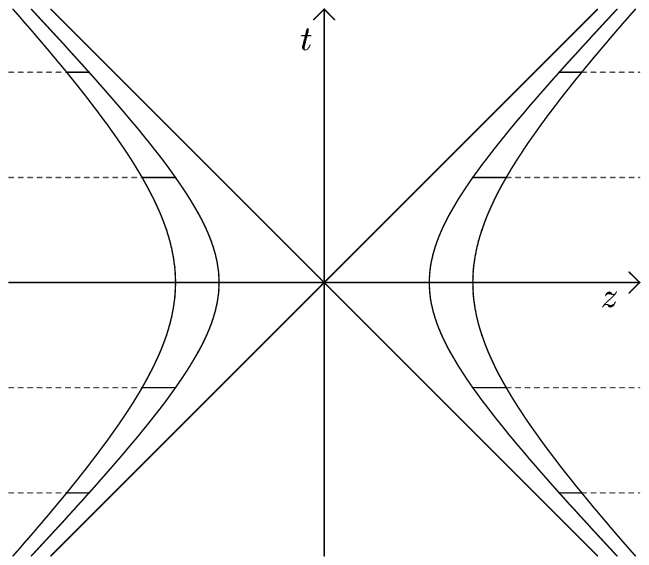,height=2in}
}
\caption{Two uniformly accelerated spinning black holes
   a) connected by a spring between them,
   b) with string extending from each of them to infinity.}
\end{figure}

\vskip 9mm

{\bf $\!\!\!\!\!\!$Acknowledgments}.
\vskip 5mm

I am very grateful to the organizers of ICGC 2000 for their hospitality.
I thank Tom\'a\v s Ledvinka for constructing the figures and help with the manuscript.
Support from the grant No. GA\v CR 202/99/0261 of the Czech Republic and
the Grant J13/98:113200004 is also gratefully acknowledged.

\vskip 4mm

\newcommand{\atque}{and }
\newcommand{\NAME}[1]{{#1}}
\let\name\NAME
\newcommand{\BY}[1]{\NAME{#1},}
\newcommand{\IN}[4]{\textit{#1}, \textbf{#2} (#3) #4}
\newcommand{\SAME}[3]{\textbf{#1} (#2) #3}
\newcommand{\TITLE}[1]{\textit{#1}}
\newcommand{\ADLIB}[1]{{#1}}
\let\Author\BY
\let\Name\BY
\newcommand{\Review}[1]{\textit{#1},}
\newcommand{\Vol}[1]{\textbf{#1}}
\newcommand{\Year}[1]{(#1)}
\newcommand{\Page}[1]{#1}

\newcommand {\ia}[1] {}
\newcommand {\TITLEA}[1] {{\it #1}}

\end{document}